\documentclass[11pt]{article}
\usepackage[utf8]{inputenc}
\usepackage{bm, commath}
\usepackage{natbib}
\usepackage[hyphens]{url}
\usepackage{caption}
\usepackage{graphicx}
\usepackage{subcaption}
\usepackage{amsmath, amsfonts, amsthm}
\usepackage{float}
\usepackage{booktabs,siunitx}
\usepackage{multirow}
\usepackage{amssymb}
\usepackage{blkarray}
\usepackage{bbm}
\usepackage{multicol}
\usepackage{authblk}
\usepackage[usenames,dvipsnames]{xcolor}
\usepackage{listings}
\usepackage{bbm}
\usepackage{textcomp}
\usepackage[margin=1in]{geometry}
\usepackage{authblk}
\usepackage[ruled]{algorithm2e}
\SetKwInput{KwParam}{Parameter}
\SetAlgoCaptionLayout{centerline}

\usepackage{sectsty}
\usepackage[onehalfspacing]{setspace}
\usepackage{float}
\usepackage{thmtools}
\declaretheorem[style=definition]{example}

\usepackage{tikz}
\usetikzlibrary{positioning,chains,fit,shapes,calc}

\newtheorem{theorem}{Theorem}
\newtheorem{definition}{Definition}
\newtheorem{corollary}{Corollary}

\newtheorem{assumption}{Assumption}
\newtheorem*{assumption*}{Assumption}

\newtheorem{proposition}{Proposition}

\newtheorem{lemma}{Lemma}

\renewcommand\thmcontinues[1]{Continued}

\usepackage{mathtools}

\makeatletter
\renewcommand{\algocf@captiontext}[2]{#1\algocf@typo. \AlCapFnt{}#2} 
\def\@algocf@capt@plain{top}
\renewcommand{\algocf@makecaption}[2]{%
  \addtolength{\hsize}{\algomargin}%
  \sbox\@tempboxa{\algocf@captiontext{#1}{#2}}%
  \ifdim\wd\@tempboxa >\hsize
  \hskip .5\algomargin%
  \parbox[t]{\hsize}{\algocf@captiontext{#1}{#2}}
  \else%
  \global\@minipagefalse%
  \hbox to\hsize{\box\@tempboxa}
  \fi%
  \addtolength{\hsize}{-\algomargin}%
}
\makeatother

\sectionfont{\bfseries\large\sffamily}%
\subsectionfont{\bfseries\sffamily\normalsize}%

\newcommand{\reviewer}[3]{
  \expandafter\newcommand\csname #1\endcsname[1]{%
    \textcolor{#3}{[\textsf{\footnotesize \textbf{#2:} ##1]}}%
  }
}

\newcommand{\btcb}{\begin{tcolorbox}}
\newcommand{\etcb}{\end{tcolorbox}}
\newcommand{\bbm}{\begin{bmatrix}}
\newcommand{\ebm}{\end{bmatrix}}
\newcommand{\bassume}{\begin{assumption}}
\newcommand{\eassume}{\end{assumption}}
\newcommand{\be}{\begin{equation}}
\newcommand{\ee}{\end{equation}}
\newcommand{\ben}{\begin{equation*}}
\newcommand{\een}{\end{equation*}}
\newcommand{\bea}{\begin{aligned}}
\newcommand{\eea}{\end{aligned}}
\newcommand{\ba}{\begin{equation}\begin{aligned}}
\newcommand{\ea}{\end{aligned}\end{equation}}
\newcommand{\bd}{\begin{definition}}
\newcommand{\ed}{\end{definition}}
\newcommand{\bprop}{\begin{proposition}}
\newcommand{\eprop}{\end{proposition}}
\newcommand{\bt}{\begin{theorem}}
\newcommand{\et}{\end{theorem}}
\newcommand{\bcor}{\begin{corollary}}
\newcommand{\ecor}{\end{corollary}}
\newcommand{\beg}{\begin{example}}
\newcommand{\eeg}{\end{example}}
\newcommand{\bnt}[1]{\begin{namedthm}{#1}}
\newcommand{\ent}{\end{namedthm}}
\newcommand{\blm}{\begin{lemma}}
\newcommand{\elm}{\end{lemma}}
\newcommand{\bp}{\begin{proof}}
\newcommand{\ep}{\end{proof}}
\newcommand{\bpb}{\begin{problem}}
\newcommand{\epb}{\end{problem}}
\newcommand{\benum}{\begin{enumerate}}
\newcommand{\eenum}{\end{enumerate}}
\newcommand{\bitem}{\begin{itemize}}
\newcommand{\eitem}{\end{itemize}}
\definecolor{firebrick}{RGB}{178,34,34}

\def\brst\begin{restatable}
\newcommand{\erst}{\end{restatable}}

\usepackage{threeparttable }
\usepackage{multicol,multirow}
\usepackage[hyphens]{url}

\reviewer{jiayao}{JZ}{NavyBlue}
\reviewer{ting}{TY}{red}
\reviewer{dan}{DR}{blue}

\title{Matching with multiple criteria and its application to health disparities research}

\author[1]{Chang Chen}
\author[2]{Zhiyu Qian}
\author[3]{Bo Zhang \thanks{Correspondence to Bo Zhang, Vaccine and Infectious Disease Divison, Fred Hutchinson Cancer Center, Seattle, Washington, 98109. Email: bzhang3@fredhutch.org}}

\affil[1]{Department of Biostatistics, University of Washington}
\affil[2]{Division of Urological Surgery, 
       Brigham and Women’s Hospital, Harvard Medical School}
\affil[3]{Vaccine and Infectious Disease Division, Fred Hutchinson Cancer Center}

\date{}
\begin{document}

\maketitle
\noindent
\textsf{{\bf Abstract}: }%
Matching is a popular nonparametric covariate adjustment strategy in empirical health services research. Matching helps construct two groups comparable in many baseline covariates but different in some key aspects under investigation. 
In health disparities research, it is desirable to understand the contributions of various modifiable factors, like income and insurance type, to the observed disparity in access to health services between different groups. 
To single out the contributions from the factors of interest, we propose a statistical matching methodology that constructs nested matched comparison groups from, for instance, White men, that resemble the target group, for instance, black men, in some selected covariates while remaining identical to the white men population before matching in the remaining covariates. Using the proposed method, we investigated the disparity gaps between white men and black men in the US in prostate-specific antigen (PSA) screening based on the 2020 Behavioral Risk Factor Surveillance System (BFRSS) database. We found a widening PSA screening rate as the white matched comparison group increasingly resembles the black men group and quantified the contribution of modifiable factors like socioeconomic status. Finally, we provide code that replicates the case study and a tutorial that enables users to design customized matched comparison groups satisfying multiple criteria.

\vspace{0.3 cm}
\section{Introduction}

\subsection{Statistical matching in observational studies}
Carefully-designed observational studies may provide a useful piece of evidence towards establishing or nullifying a scientifically meaningful association or causal conclusion. Observational studies often suffer from both overt and hidden bias due to systematic differences in pretreatment covariates between two comparison groups \citep{rosenbaum2002observe}. The bias may lead to researchers concluding a spurious association or a cause-and-effect relationship after conducting a na\"ive outcome analysis comparing two groups. Statistical matching is a widely used nonparametric approach to reducing the \emph{overt} bias by adjusting for \emph{observed} differences in covariates between comparison groups. The overall guiding principle of statistical matching is to construct a treatment group and a comparison group, or two comparison groups in general, that are well-balanced in key observed covariates. Matching is particularly useful in some contexts. First, when two comparison groups are poorly overlapped, model-based methods (including methods based on modeling potential outcomes, weighting, or a combination of both) could rely heavily on model extrapolation (see, e.g., \citealp{Kang2007demystifying}); in these cases, statistical matching, as a pre-processing step, could help researchers focus on the well-overlapped covariate space and minimize the risk of model misspecification and over-extrapolation \citep{rubin1979using,ho2007matching}. Second, modern statistical matching could be particularly useful when the goal is to control for observed covariates with many categories, for instance, extensive medical history and comorbidities often encountered in large healthcare databases \citep{rosenbaum2020modern}. Importantly, like any covariate adjustment method, statistical matching cannot adjust for hidden bias and a sensitivity analysis is essential in observational studies to examine the level of unmeasured confounding needed to nullify the observed association (see, e.g., \citealp{rosenbaum2002observe, rosenbaum2010design,vanderweele2017sensitivity,zhang2020calibrated,zhang2022semi}, among many others). 

\subsection{Usefulness of multiple comparisons in health disparities research}
\label{subsec: multiple matched comparisons}
In health disparities research, it is of interest to understand the contributions of modifiable factors, like income and insurance, to the observed disparities in accessing health services between different racial/ethnic groups (e.g., Black and White people). In these circumstances, multiple matched comparisons that sequentially adjust for baseline covariates could be useful (see, e.g., \citealp{silber2013characteristics, silber2014racial, nogueira2022association}). One example of an application of this strategy is \citeauthor{nogueira2022association}'s \citeyearpar{nogueira2022association} analysis of the association between race and receipt of proton beam therapy (PBT) among patients with newly diagnosed cancer in the US. \citet{nogueira2022association} considered multiple matched comparisons. In an unadjusted comparison between Black and White patients, \citet{nogueira2022association} observed a sharp difference in the baseline characteristics, including age group, sex, cancer site, cancer stage, socioeconomic status, and insurance type, among others. In a first propensity-score-based matched comparison $\textsf{M1}$, \citet{nogueira2022association} matched on variables related to biology (age and sex), pathology (cancer site and cancer stage) and availability of the therapy (region). Presumably, in their first matched comparison $\textsf{M1}$, by matching only on variables related to biology, pathology and availability of the therapy, \citet{nogueira2022association} was trying to examine how much difference in the PBT receipt in the unadjusted analysis could be attributed to these factors \emph{alone}. In a second matched comparison, \citet{nogueira2022association} matched on patients’ quintile of median income by zip codes, in addition to variables in the first matched comparison, to further examine the contribution of the income gap to the persisting disparity in the first matched comparison. \citet{nogueira2022association} found that the racial disparities narrowed but still remained significant after further matching on income.
This approach is known as \emph{tapered matching} in the statistical matching literature
(see, e.g., \citealp{daniel2008algorithm, rosenbaum2013using, yu2021information}). Tapered matching constructs multiple matched comparison groups that increasingly resemble  the treated group by controlling for additional covariates and is intended to answer various research questions about the root causes of the observed outcome difference in two comparison groups.

\subsection{Our contribution}
\label{subsec: intro contribution}
We argue that, in order to best understand the contributions of various factors to the observed disparity, it is helpful to create a matched comparison group (in this case white patients) that satisfies two criteria. For instance, in the matched comparison $\textsf{M1}$ discussed in the last section, it is helpful that the matched white patients resemble (1) black patients in biology, pathology and availability of the therapy and (2) the population of white patients before matching in other aspects like income and insurance type. By comparing the (possibly persisting) disparity observed in this matched comparison to that in the unadjusted analysis, researchers could better single out the contribution of biology, pathology and availability of the therapy to the disparity observed in an unadjusted analysis. Similarly, for the matched comparison $\textsf{M2}$, it is helpful that the matched white patients resemble the black patients in biology, pathology, availability of the therapy, and income, while remaining identical or near-identical to the white patients population before matching in insurance type. In this way, a comparison of the results obtained from $\textsf{M2}$ to $\textsf{M1}$ helps illuminate the contribution of income to the persisting disparity observed in $\textsf{M1}$, because two white comparison groups in $\textsf{M1}$ and $\textsf{M2}$ are now near-identical in all but income.

How does this strategy improve upon what tapered matching has already been doing? In their matched comparison $\textsf{M2}$ using tapered matching, although \citet{nogueira2022association} chose deliberately not to match on insurance, the final matched white patients still became more similar to the black patients in the insurance type. For instance, the proportion of patients on Medicare is 38.9\% among black patients, 46.5\% among all white patients, and 40.0\% among white patients in $\textsf{M2}$. A comparison of the disparity observed in $\textsf{M2}$ and that in $\textsf{M1}$ then failed to single out the sole contribution of income, but rather reported the contribution from a mixture of two factors: income and, to some extent, insurance type. 

To summarize, our goal is to construct multiple, nested matched comparisons as follows. Starting from an unadjusted group of white patients in the population, the first match would construct a matched comparison group that resembles the black patients in covariates $(X_1, \dots, X_{I_1})$ while remaining the same as the white population in covariates $(X_{I_1 + 1}, \dots, X_{K})$. In the second match, the updated comparison group would resemble the black patients in the covariates $(X_1, \dots, X_{I_1 + I_2})$ while remaining the same as the white population in the other covariates $(X_{I_1 + I_2 + 1}, \dots, X_{K})$. The process could continue until reaching the final matched comparison where the constructed white patients comparison group resembles the black patients in all $K$ covariates deemed relevant in the analysis. 

\subsection{Organization of the article}
\label{subsec: organization}
The rest of the article is organized as follows. Section \ref{sec: matching in bipartite graph} provides a brief overview of constructing a matched comparison group using a network-flow-based optimization algorithm built upon a bipartite network. Section \ref{sec: tripartite method} describes a matching framework built around a tripartite graph and adapts it to deliver our proposed design for health disparities research. Section \ref{sec: case study} applies the method to investigating the disparities gap in prostate cancer screening in the US.  Section \ref{sec: code} introduces how to implement the proposed method in a modularized manner using the \textsf{R} package \textsf{match2C}. Section \ref{sec: discussion} concludes with a discussion.

\section{Optimal matching based on a bipartite graph}
\label{sec: matching in bipartite graph}
A conventional greedy match algorithm constructs two matched groups comparable in many baseline covariates via a stepwise decision-making procedure, where each decision is best among the choices that are available, and it typically yields a sub-optimal match. In a seminal paper, \citet{rosenbaum1989optimal} first bridged the statistical literature and operations research literature and formulated the statistical matching problem as a combinatorial optimization problem of finding a minimum-cost network flow in a bipartite graph.

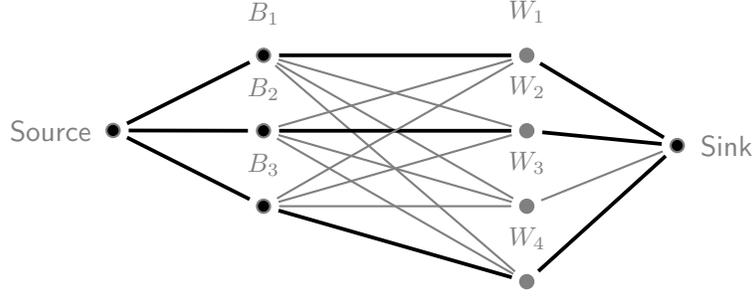
\begin{figure}[ht]
\centering
\begin{tikzpicture}[thick, color = gray,
  every node/.style={draw,circle},
  fsnode/.style={fill=black, inner sep = 0pt, minimum size = 5pt},
  ssnode/.style={fill=gray, inner sep = 0pt, minimum size = 5pt},
  every fit/.style={ellipse,draw,inner sep=-2pt,text width=2cm},
  shorten >= 3pt,shorten <= 3pt
]


\begin{scope}[start chain=going below,node distance=8mm]
\foreach \i in {1,2,3}
  \node[fsnode,on chain] (t\i) [label=above: {\small$B_\i$} ] {};
\end{scope}

\begin{scope}[xshift=3.5cm,yshift=0cm,start chain=going below,node distance=8mm]
\foreach \i in {1,2,3,4}
  \node[ssnode,on chain] (c\i) [label=above: {\small$W_\i$}] {};
\end{scope}

\node [fill = black, inner sep = 0pt, minimum size = 5pt, label=left: \textsf{Source}] at (-2, -1) (source) {};

\node [fill = black, inner sep = 0pt, minimum size = 5pt, label=right: \textsf{Sink}] at (5.5, -1.2) (sink) {};


\draw [line width = 0.5mm, color = black] (t1) -- (c1);
\draw (t1) -- (c2);
\draw (t1) -- (c3);
\draw (t1) -- (c4);

\draw (t2) -- (c1);
\draw [line width = 0.5mm, color = black] (t2) -- (c2);
\draw (t2) -- (c3);
\draw (t2) -- (c4);

\draw (t3) -- (c1);
\draw (t3) -- (c2);
\draw (t3) -- (c3);
\draw [line width = 0.5mm, color = black](t3) -- (c4);

\draw [line width = 0.5mm, color = black] (source) -- (t1);
\draw [line width = 0.5mm, color = black] (source) -- (t2);
\draw [line width = 0.5mm, color = black] (source) -- (t3);

\draw [line width = 0.5mm, color = black] (c1) -- (sink);
\draw [line width = 0.5mm, color = black] (c2) -- (sink);
\draw (c3) -- (sink);
\draw [line width = 0.5mm, color = black](c4) -- (sink);

\end{tikzpicture}
\caption{Network-flow representation of a basic matching example. Four white patients $W_1$, $W_2$, $W_3$ and $W_4$ are to be matched to three black men patients $B_1$, $B_2$, and $B_3$. Each edge connecting $B_i$, $i = 1,2,3$, and $W_j$, $j = 1,2,3,4$, is associated with a flow capacity (equal to $1$ in pair matching) and a cost, $\delta_{i,j}$, equal to a pre-specified covariate distance between $B_i$ and $W_j$. The bold black lines correspond to three optimal matched pairs: $(B_1, W_1)$, $(B_2, W_2)$, and $(B_3, W_4)$. }
\label{fig: bipartite plot}
\end{figure}

We illustrate the advantage of a network-flow-based method by comparing it to the greedy algorithm in a toy example. Figure \ref{fig: bipartite plot} represents three black men patients, {$B_1$, $B_2$, $B_3$}, and four white men patients, {$W_1$, $W_2$, $W_3$, $W_4$}, as vertices in a graph.
The pre-specified covariate distance, denoted as $\delta_{i,j}$, represents a measure of similarity between each pair of black ($B_i$) and white ($W_j$) patients, and it is associated with the edge $e_{i,j}$ connecting $B_i$ and $W_j$. 
Within-pair distances $\delta_{i,j}$  are as follows:
$\{
\delta_{1,1} = 1,
\delta_{1,2} = 0,
\delta_{1,3} = 2,
\delta_{1,4} = 3,
\delta_{2,1} = 3,
\delta_{2,2} = 1,
\delta_{2,3} = 4,
\delta_{2,4} = 5,
\delta_{3,1} = 2,
\delta_{3,2} = 2,
\delta_{3,3} = 2,
\delta_{3,4} = 1
\}$. In the network flow optimization setting, we send three units of ``flows" out from the ``Source" vertex to the ``Sink" vertex. Each flow can only bypass one of the above-defined edges. Each of the three emitting edges and four sinking edges is assigned a cost or distance of 0. The cost between any pair of a black patient vertex and a white patient vertex is equal to a pre-specified covariate distance, $\delta_{i,j}$. One can check that the flow highlighted in black in Figure \ref{fig: bipartite plot}, $\{e_{1,1},e_{2,2},e_{3,4}\}$, has a total distance of $3$ and is a minimum cost flow that minimizes the total covariate distance within the pair-matched set. There are several forms of within-matched set covariate distances.
Some common choices include the $L_1$ distance in the estimated propensity score, Mahalanobis distance between multivariate covariates, and a robust version of it. A robust Mahalanobis distance is sometimes preferred as it obtains robustness to outliers by using rank-adjusted covariate coordinates \citep{rosenbaum2010design}. 



On the other hand, according to a greedy algorithm, one would start with a match of minimum distance, $e_{1,2}$ in this toy example, and remove these two vertices ($B_1$ and $W_2$) from further consideration. This process is then repeated until all black patients represented by $(B_1, B_2, B_3)$ have been paired. In this example, the greedy algorithm finally selects $\{e_{1,2},e_{3,4},e_{2,2}\}$ with a total distance of $4$ ans is inferior to the match produced by the network-flow-based algorithm.

\section{From bipartite to tripartite}
\label{sec: tripartite method}

\subsection{Generic network structure}\label{subsec: tripartite structure}
More recently, \citet{zhang2021matching} proposed a minimum-cost network flow algorithm built upon a tripartite graph for multivariate matching. The network infrastructure of a generic tripartite network is depicted in Figure \ref{fig: tripartite network} and can be formalized as follows. A directed graph $\mathcal{G}$ is a set of vertices $\mathcal{V}$ and a set of directed edges $\mathcal{E}$ consisting of ordered pairs $e = (v_1, v_2)$ of distinct vertices. A total of $L_1$ units, $\mathcal{L}_1 = \{\tau_1, \dots, \tau_{L_1}\}$, are represented in the far left part of the network by nodes labeled $\tau_{l_1}$, $l_1 = 1, \dots, L_1$. In parallel, a total of $L_3$ units, $\mathcal{L}_3 = \{\omega_1, \dots, \omega_{L_3}\}$, are represented in the far right part of the network by nodes labeled $\omega_{l_3}$, $l_3 = 1, \dots, L_3$. Finally, a total of $L_2$ units, $\mathcal{L}_2 = \{\gamma_1, \dots, \gamma_{L_2}\}$, occur twice in the network and are represented by nodes labeled $\gamma_{l_2}$ and their mirror copies $\overline\gamma_{l_2}$, ${l_2} = 1, \dots, L_2$. When adopting this tripartite network structure in a statistical matching problem, as opposed to a standard bipartite network structure depicted in Figure \ref{fig: bipartite plot}, the goal is to select a subset of units $\mathcal{L}'_2$ from $\mathcal{L}_2$ so that $\mathcal{L}'_2$ resemble units in both $\mathcal{L}_1$ and $\mathcal{L}_3$, but possibly in different aspects. Write $|S|$ for the number of elements of a finite set $S$, so $|\mathcal{L}_1|=L_1$, $|\mathcal{L}_3|=L_3$ and $|\mathcal{L}_2|=L_2$. We require that $L_2 > \max\{L_1, L_3\}$ and that $L_2 > \text{LCM}(L_1, L_3)$, where LCM denotes the least common multiple of $L_1$ and $L_3$.

There is a source vertex $\xi$ and a sink vertex $\overline{\xi}$, so that the network consists of $|\mathcal{V}| = L_1 + 2L_2 + L_3 + 2$ total vertices:
\begin{equation*}
    \mathcal{V} = \left\{\xi, \tau_1, \dots, \tau_{L_1}, \gamma_1, \dots, \gamma_{L_2}, \overline\gamma_1, \dots, \overline\gamma_{L_2}, \omega_1, \dots, \omega_{L_3},  \overline\xi\right\}.
\end{equation*}
This tripartite network consists of the following edges:
\begin{equation*}
    \mathcal{E} = \left\{(\xi, \tau_{l_1}), (\tau_{l_1}, \gamma_{l_2}), (\gamma_{l_2}, \overline\gamma_{l_2}), (\overline\gamma_{l_2}, \omega_{l_3}), (\omega_{l_3}, \overline\xi), ~l_1 = 1, \dots, L_1,~l_2 = 1, \dots, L_2, ~l_3 = 1, \dots, L_3 \right\}.
\end{equation*}
Thus, there are a total of $|\mathcal{E}| = L_1 + L_2 + L_3 + L_1L_2 + L_2L_3$ edges.

Let $\text{cap}(e)\geq 0$ denote the capacity of an edge $e \in \mathcal{E}$. In the basic network structure depicted in Figure \ref{fig: tripartite network}, all edges of the form $e(\xi, \tau_{l_1})$, ${l_1} = 1, \dots, L_1$, have capacity $\kappa = \text{LCM($L_1$, $L_3$)}/L_1$. Analogously, all edges of the form $e(\omega_{l_3}, \overline{\xi})$, $l_3 = 1, \dots, L_3$, have capacity $\kappa' = \text{LCM($L_1$, $L_3$)}/L_3$. All other edges, including those of the form $e(\tau_{l_1}, \gamma_{l_2})$, $e(\gamma_{l_2}, \overline\gamma_{l_2})$, and $e(\overline\gamma_{l_2}, \omega_{l_3})$, $l_1 = 1 \dots, L_1$, $l_3 = 1, \dots, L_3$, and $l_2 = 1, \dots, L_2$, all have capacity $1$. 

A feasible flow $f(\cdot)$ is a function that assigns a nonnegative integer to each edge $e \in \mathcal{E}$, such that
\begin{itemize}
    \item[(i)] $0 \leq f(e) \leq \text{cap}(e),~ \text{for } e\in\mathcal{E}$;
    \item[(ii)] All flow units supplied at $\xi$ are absorbed at $\overline\xi$, i.e., $\sum_{l_1 = 1}^{L_1} f\{(\xi, \tau_{l_1})\} = \sum_{l_3 = 1}^{L_3} f\{(\omega_{l_3}, \overline\xi)\} = \text{LCM}(L_1, L_3)$;
    \item[(iii)] The flow into each vertex equals the flow out from each vertex, i.e., write $\mathcal{E}^\prime \subset \mathcal{E}$ for the set of edges that include neither the source $\xi$ nor the sink $\overline\xi$, then $\sum_{(a,b)\in\mathcal{E}^\prime} f\{(a,b)\} = \sum_{(b,c)\in\mathcal{E}^\prime} f\{(b,c)\}$ for all $b \in \mathcal{V}\backslash\{\xi, \overline\xi\}$.
\end{itemize}

A feasible flow produces a feasible match. Because all edges of the form $e(\tau_{l_1}, \gamma_{l_2})$ and $e(\overline\gamma_{l_2}, \omega_{l_3})$ have capacity $1$, quantities of the form $f(\tau_{l_1}, \gamma_{l_2})$ and $f(\overline\gamma_{l_2}, \omega_{l_3})$ all take an integer value of either $0$ or $1$. Formally, the matched sample $\mathcal{M}(f)$ is defined by
\begin{equation*}
    \mathcal{M}(f) = \left\{(\tau_{l_1}, \gamma_{l_2}), (\overline\gamma_{l_2}, \omega_{l_3})\in \mathcal{E}: f\{(\tau_{l_1}, \gamma_{l_2})\} = f(\overline\gamma_{l_2}, \omega_{l_3})= 1\right\}.
\end{equation*}
Lastly, $\mathfrak{M} = \left\{ \mathcal{M}(f): \text{for some feasible } f \right\}$ denotes the set of all feasible matched samples. According to the tripartite network and edges' associated capacities in the network, we are constructing one matched sample via conducting two optimal bipartite matching at one shot: a $1$-to-$\kappa$-match facilitated by the left part of the network and a $1$-to-$\kappa'$-match facilitated by the right part of the network.

\subsection{Cost and optimal matched sample}
\label{subsec: cost}
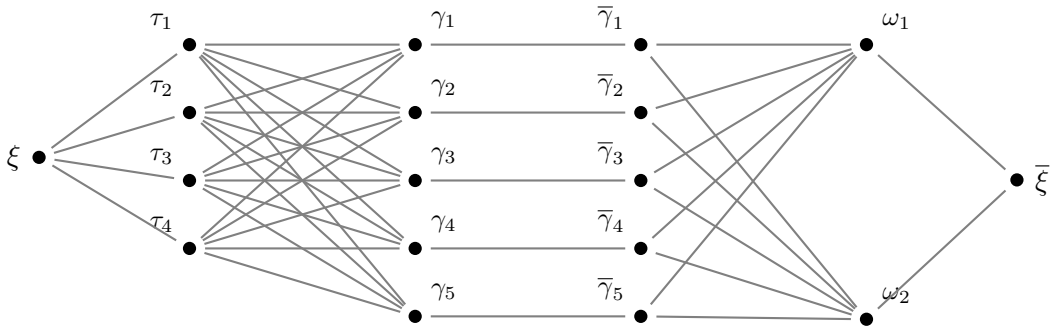
\begin{figure}[ht]
\centering
\begin{tikzpicture}[thick, color = black,
  fsnode/.style={circle, fill=black, inner sep = 0pt, minimum size = 5pt},
  ssnode/.style={circle, fill=black, inner sep = 0pt, minimum size = 5pt},
  shorten >= 3pt,shorten <= 3pt
]


\begin{scope}[start chain=going below,node distance=7mm]
\foreach \i in {1,2,3,4}
  \node[fsnode,on chain] (r\i) [label=above left: {\small$\tau_\i$} ] {};
\end{scope}

\begin{scope}[xshift=3cm,yshift=0cm,start chain=going below,node distance=7mm]
\foreach \i in {1,2,3,4,5}
  \node[ssnode,on chain] (t\i) [label=above right: {\small$\gamma_\i$}] {};
\end{scope}

\begin{scope}[xshift=6cm,yshift=0cm,start chain=going below,node distance=7mm]
\foreach \i in {1,2,3,4,5}
  \node[ssnode,on chain] (tt\i) [label=above left: {\small$\overline\gamma_\i$}] {};
\end{scope}

\begin{scope}[xshift=9cm,yshift=0cm,start chain=going below,node distance=34.5mm]
\foreach \i in {1,2}
  \node[ssnode,on chain] (c\i) [label=above right: {\small$\omega_\i$}] {};
\end{scope}

\node [circle, fill = black, inner sep = 0pt, minimum size = 5pt, label=left: $\xi$] at (-2, -1.5) (source) {};

\node [circle, fill = black, inner sep = 0pt, minimum size = 5pt, label=right: $\overline\xi$ ] at (11, -1.8) (sink) {};



\foreach \i in {1,2,3,4} {
   \draw[color=gray] (source) -- (r\i);
   }

\foreach \i in {1,2,3,4} {
   \foreach \j in {1,2,3,4,5} {
   \draw[color=gray] (r\i) -- (t\j);
   }
 } 
 
\foreach \i in {1,2,3,4,5} {
   \draw[color=gray] (t\i) -- (tt\i);
}  
 
\foreach \i in {1,2,3,4,5} {
   \foreach \j in {1,2} {
   \draw[color=gray] (tt\i) -- (c\j);
   }
} 

\foreach \i in {1,2} {
   \draw[color=gray] (c\i) -- (sink);
}


\end{tikzpicture}
\caption{A basic illustrative plot of a generic tripartite network for matching with two criteria.
consisting of four units and five potential control units.}
\label{fig: tripartite network}
\end{figure}

The cost of each edge helps select the matched sample that satisfies the desired features of the problem. Let $\text{cost}(e)$ denote the cost associated with edge $e \in \mathcal{E}$. In the basic network structure depicted in Figure \ref{fig: tripartite network}, we let $\text{cost}\{(\xi, \tau_{l_1})\} = \text{cost}\{(\gamma_{l_2}, \overline\gamma_{l_2})\} = \text{cost}\{ \omega_{l_3}, \overline\xi\} = 0$ for $l_1 = 1, \dots, L_1$, $L_3 = 1, \dots, L_3$ and $l_2 = 1, \dots, L_2$. Each edge of the form $e(\tau_{l_1}, \gamma_{l_2})$ is associated with a cost $\delta_{\tau_{l_1}, \gamma_{l_2}} \geq 0$. We equate $\delta_{\tau_{l_1}, \gamma_{l_2}}$ to a measure of the similarity between unit $\tau_{l_1}$ and unit $\gamma_{l_2}$, that is,
\begin{align*}
    \text{cost}\{e(\tau_{l_1}, \gamma_{l_2})\} &= \delta_{\tau_{l_1}, \gamma_{l_2}}, ~\forall~e(\tau_{l_1}, \gamma_{l_2}) \in \mathcal{E}.
\end{align*}
Analogously, each edge of the form $e(\overline\gamma_{l_2}, \omega_{l_3})$ is associated with a cost $\Delta_{\overline\gamma_{l_2}, \omega_{l_3}} \geq 0$ which is equal to a measure of the similarity between unit $\omega_{l_3}$ and unit $\gamma_{l_2}$, that is,
\begin{align*}
    \text{cost}\{e(\overline\gamma_{l_2}, \omega_{l_3})\} &= \Delta_{\overline\gamma_{l_2}, \omega_{l_3}}, ~\forall~e(\overline\gamma_{l_2}, \omega_{l_3}) \in \mathcal{E}.
\end{align*}
Lastly, we let a parameter $\lambda \geq 0$ control the relative importance of the two costs, so that the cost of a feasible flow $f$ in the network becomes:
\begin{equation}\label{eq: tripartite flow cost}
    \text{cost}(f) = \sum_{(\tau_{l_1}, \gamma_{l_2}) \in \mathcal{M}(f)}\delta_{\tau_{l_1}, \gamma_{l_2}} + 
    \lambda \Big\{ \sum_{(\overline\gamma_{l_2}, \omega_{l_3}) \in \mathcal{M}(f)}\Delta_{\overline\gamma_{l_2}, \omega_{l_3}} \Big\}.
\end{equation}
A minimum cost flow is a feasible flow $g(\cdot)$ such that $\text{cost}(g) \leq \text{cost}(f)$ for every feasible flow $f(\cdot)$.

\subsection{Adapting the tripartite network to health disparities research}
\label{subsec: tripartite in disparity research}
As discussed extensively in Section \ref{subsec: multiple matched comparisons}, our goal is to construct a cohort of white patients who satisfy the following two criteria: (i) selected white patients resemble black patients in covariates $\boldsymbol{X}$; and (ii) selected white patients resemble white patients before matching in covariates $\widetilde{\boldsymbol{X}}$, where $\boldsymbol{X} \bigcap \widetilde{\boldsymbol{X}} = \emptyset$. Let $\mathcal{B}$ denote the set of black patients and $\mathcal{W}$ the set of white patients before matching. Let $\mathcal{W}'$ be a random sample of size $|\mathcal{B}|$ from $\mathcal{W}$. For instance, in the case study to be discussed in Section \ref{sec: case study}, $\mathcal{B}$ consists of $2,507$ black people, $\mathcal{W}$ consists of $24,344$ white people, and $\mathcal{W}'$ consists of a random sample of size $2,507$ from $\mathcal{W}$. To select the desired matched comparison group using the tripartite network structure described in Section \ref{subsec: tripartite structure}, we let $\mathcal{L}_1 = \mathcal{B}$, $\mathcal{L}_2 = \mathcal{W}$, and $\mathcal{L}_3 = \mathcal{W}'$. We further let $\delta(\tau_{l_1}, \gamma_{l_2})$ equal a measure of the similarity between units in $\mathcal{W}$ and $\mathcal{B}$ in covariates $\boldsymbol{X}$ and let $\Delta(\overline\gamma_{l_2}, \omega_{l_3})$ equal a measure of the similarity between units in $\mathcal{W}$ and $\mathcal{W}'$ in covariates $\widetilde{\boldsymbol{X}}$. For instance, one may equal $\delta(\tau_{l_1}, \gamma_{l_2})$ to the absolute difference in \citeauthor{rosenbaum1983central}'s \citeyearpar{rosenbaum1983central} propensity score estimated using data $\mathcal{B}$ and $\mathcal{W}$ and based on covariates $\boldsymbol{X}$ alone. Alternatively, one may let $\delta(\tau_{l_1}, \gamma_{l_2})$ represent the (robust) Mahalanobis distance in $\boldsymbol{X}$ or the (robust) Mahalanobis distance within a propensity score caliper. Analogously, one may let $\Delta(\overline\gamma_{l_2}, \omega_{l_3})$ equal (1) the propensity score estimated using data $\mathcal{W}$ and $\mathcal{W}'$ and based on covariates $\widetilde{\boldsymbol{X}}$ alone; (2) (robust) Mahalanobis distance based on covariates $\widetilde{\boldsymbol{X}}$ alone; or (3) a combination of both. According to this framework, in an ideal situation, the selected matched comparison group will consist of white people who remain similar to the white population in aspects captured by $\widetilde{\boldsymbol{X}}$ while resembling the black people in aspects captured by $\boldsymbol{X}$.

\subsection{Connection to previous works}
\label{subsec: connection to previous works}
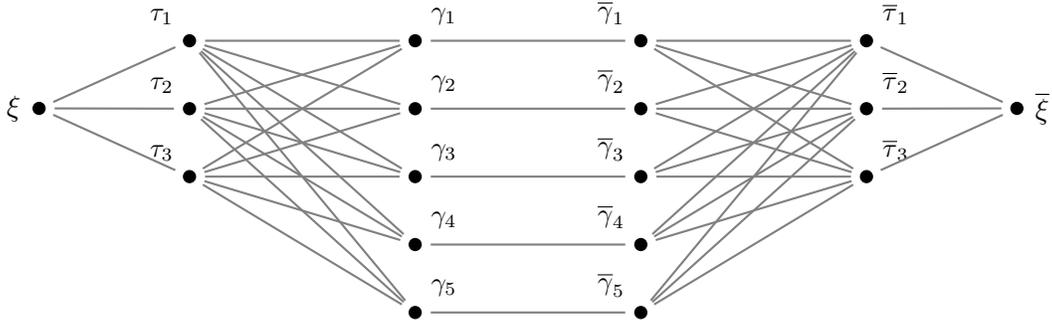
\begin{figure}[ht]
\centering
\begin{tikzpicture}[thick, color = black,
  fsnode/.style={circle, fill=black, inner sep = 0pt, minimum size = 5pt},
  ssnode/.style={circle, fill=black, inner sep = 0pt, minimum size = 5pt},
  shorten >= 3pt,shorten <= 3pt
]


\begin{scope}[start chain=going below,node distance=7mm]
\foreach \i in {1,2,3}
  \node[fsnode,on chain] (r\i) [label=above left: {\small$\tau_\i$} ] {};
\end{scope}

\begin{scope}[xshift=3cm,yshift=0cm,start chain=going below,node distance=7mm]
\foreach \i in {1,2,3,4,5}
  \node[ssnode,on chain] (t\i) [label=above right: {\small$\gamma_\i$}] {};
\end{scope}

\begin{scope}[xshift=6cm,yshift=0cm,start chain=going below,node distance=7mm]
\foreach \i in {1,2,3,4,5}
  \node[ssnode,on chain] (tt\i) [label=above left: {\small$\overline\gamma_\i$}] {};
\end{scope}

\begin{scope}[xshift=9cm,yshift=0cm,start chain=going below,node distance=7mm]
\foreach \i in {1,2,3}
  \node[ssnode,on chain] (c\i) [label=above right: {\small$\overline\tau_\i$}] {};
\end{scope}

\node [circle, fill = black, inner sep = 0pt, minimum size = 5pt, label=left: $\xi$] at (-2, -0.9) (source) {};

\node [circle, fill = black, inner sep = 0pt, minimum size = 5pt, label=right: $\overline\xi$ ] at (11, -0.9) (sink) {};



\foreach \i in {1,2,3} {
   \draw[color=gray] (source) -- (r\i);
   }

\foreach \i in {1,2,3} {
   \foreach \j in {1,2,3,4,5} {
   \draw[color=gray] (r\i) -- (t\j);
   }
 } 
 
\foreach \i in {1,2,3,4,5} {
   \draw[color=gray] (t\i) -- (tt\i);
}  
 
\foreach \i in {1,2,3,4,5} {
   \foreach \j in {1,2,3} {
   \draw[color=gray] (tt\i) -- (c\j);
   }
} 

\foreach \i in {1,2,3} {
   \draw[color=gray] (c\i) -- (sink);
}

\end{tikzpicture}
\caption{A basic illustrative plot of a tripartite network consisting of three treated units and five potential control units.}
\label{fig: fine balance network}
\end{figure}

A tripartite network structure has been leveraged in the literature in two distinct contexts. \citet{zhang2021matching} proposed to use a tripartite network structure to construct matched pairs (or matched sets in general) from the set of treated units $\mathcal{T}$ and control units $\mathcal{C}$ that are (i) closely matched in key covariates; and (ii) balanced in high-dimensional covariates including the propensity score distributions. To achieve this using the network infrastructure described in Section \ref{subsec: tripartite structure}, simply let $\mathcal{L}_1 = \mathcal{L}_3 = \mathcal{T}$, so that vertices $\omega_{l_3}$ in Figure \ref{fig: tripartite network} become mirror copies of treated units and are now denoted as $\overline\tau$ instead. In other words, treated units appear twice in the network, in the far left and far right, and the control units are sandwiched between them. Figure \ref{fig: fine balance network} illustrates the network where the sample to be matched consists of $3$ treated units and $5$ control units. \citet{zhang2021matching} advocated equating $\delta(\tau_t, \gamma_c)$ to a measure of covariate similarity like the Mahalanobis distance, or a robust version of it, and designing $\Delta(\tau_t, \gamma_c)$ in a way that minimizes the Earth Mover's distance between the estimated propensity score distributions of the treated group and the matched control group. We direct interested readers to \citet{zhang2021matching} for additional details. Under this framework, the final treated-to-control matched pairs $(\tau_t, \gamma_c)$ can be obtained from flow $f$ as follows:
\begin{equation*}
    \mathcal{M}(f) = \left\{(\tau_t, \gamma_c), (\overline\gamma_c, \overline\tau_t)\in \mathcal{E}: f\{(\tau_t, \gamma_c)\} = f(\overline\gamma_c, \overline\tau_t)= 1\right\}.
\end{equation*}

\begin{figure}[h]
\centering
\begin{tikzpicture}[thick, color = black,
  fsnode/.style={circle, fill=black, inner sep = 0pt, minimum size = 5pt},
  ssnode/.style={circle, fill=black, inner sep = 0pt, minimum size = 5pt},
  shorten >= 3pt,shorten <= 3pt
]


\begin{scope}[start chain=going below,node distance=7mm]
\foreach \i in {1,2}
  \node[fsnode,on chain] (r\i) [label=above left: {\small$\kappa_\i$} ] {};
\end{scope}

\begin{scope}[xshift=3cm,yshift=0cm,start chain=going below,node distance=7mm]
\foreach \i in {1,2,3}
  \node[ssnode,on chain] (t\i) [label=above right: {\small$\tau_\i$}] {};
\end{scope}

\begin{scope}[xshift=6cm,yshift=0cm,start chain=going below,node distance=7mm]
\foreach \i in {1,2,3}
  \node[ssnode,on chain] (tt\i) [label=above left: {\small$\overline\tau_\i$}] {};
\end{scope}

\begin{scope}[xshift=9cm,yshift=0cm,start chain=going below,node distance=7mm]
\foreach \i in {1,2,3,4,5}
  \node[ssnode,on chain] (c\i) [label=above right: {\small$\gamma_\i$}] {};
\end{scope}

\node [circle, fill = black, inner sep = 0pt, minimum size = 5pt, label=left: $\xi$] at (-2, -0.5) (source) {};

\node [circle, fill = black, inner sep = 0pt, minimum size = 5pt, label=right: $\overline\xi$ ] at (11, -1.9) (sink) {};



\foreach \i in {1,2} {
   \draw[color=black, line width = 0.5mm] (source) -- (r\i);
   }

\foreach \i in {1,2} {
   \foreach \j in {1,2,3} {
   \draw[color=gray] (r\i) -- (t\j);
   }
 } 

\draw[color = black, line width = 0.5mm] (r1) -- (t3);
\draw[color = black, line width = 0.5mm] (r2) -- (t1);

\foreach \i in {1,2,3} {
   \draw[color=gray] (t\i) -- (tt\i);
}

\draw[color = black, line width = 0.5mm] (t3) -- (tt3);
\draw[color = black, line width = 0.5mm] (t1) -- (tt1);
 
\foreach \i in {1,2,3} {
   \foreach \j in {1,2,3,4,5} {
   \draw[color=gray] (tt\i) -- (c\j);
   }
} 
\draw[color = black, line width = 0.5mm] (tt3) -- (c2);
\draw[color = black, line width = 0.5mm] (tt1) -- (c4);

\foreach \i in {1,2,3,4,5} {
   \draw[color=gray] (c\i) -- (sink);
}

\draw[color = black, line width = 0.5mm] (c2) -- (sink);
\draw[color = black, line width = 0.5mm] (c4) -- (sink);

\node[text width=2cm, align = center] at (0,1.5) {\small\textsf{Template}};
\node[text width=2cm, align = center] at (3,1.5) {\small\textsf{Treated}};
\node[text width=2cm, align = center] at (6,1.5) {\small\textsf{Treated}};
\node[text width=2cm, align = center] at (9,1.5) {\small\textsf{Control}};

\end{tikzpicture}
\caption{A basic illustrative network of a tripartite pair matching consisting of two template units, three treated units and five potential control units. The bold black lines correspond to two optimal matched pairs: ($\tau_1$, $\gamma_4$), ($\tau_3$, $\gamma_2$).}
\label{fig: template matching network}
\end{figure}
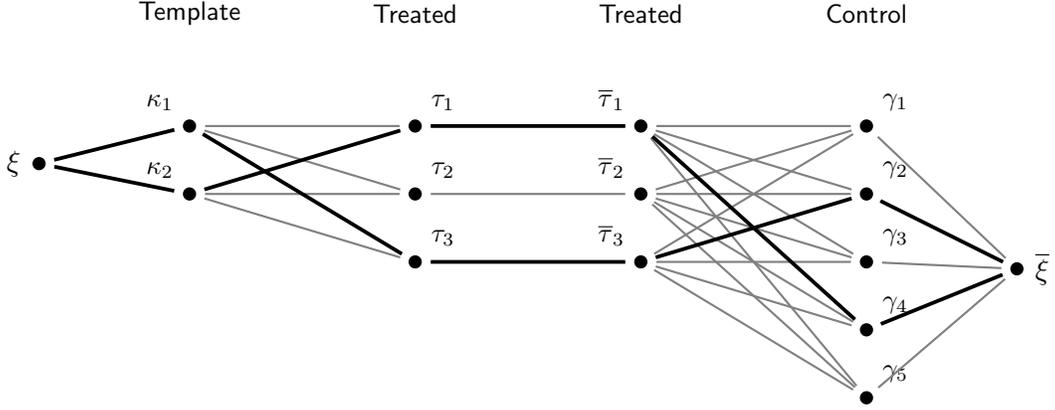

More recently, \citet{zhang2022towards} used a tripartite network to construct closely matched pairs from treated units $\mathcal{T}$ and control units $\mathcal{C}$ that further resemble a template $\mathcal{R}$ from a target population. To achieve this, let $\mathcal{L}_1 = \mathcal{R}$, $\mathcal{L}_2 = \mathcal{T}$ and $\mathcal{L}_3 = \mathcal{C}$. See Figure \ref{fig: template matching network} for an example where there are $3$ treated units and $5$ control units, and researchers further have a template consisting of $2$ units. Under this framework, costs associated with edges of the form $e(\kappa_r, \tau_t)$ facilitate selecting treated units that resembling the template while costs associated with edges of the form $e(\overline\tau_t, \gamma_c)$ facilitate the actual pair matching between treated and control units. \citet{zhang2022towards} advocated adding a large penalty to the right part of the network so that the internal validity of the matched sample is maximally maintained before pursuing its generalizability towards the target population. In Figure \ref{fig: template matching network}, a pair-matched sample $\mathcal{M}(f)$ is defined as:
\begin{equation*}
    \mathcal{M}(f) = 
    \Big \{
    (\tau_t, \gamma_c) \in \mathcal{E}: f\{(\overline\tau_t, \gamma_c)\} = 1 \Big \}.
\end{equation*}


\section{Case study: disparity gap in prostate cancer screening}
\label{sec: case study}
\subsection{Is prostate-specific antigen screening rate higher in younger Black men compared to White men?}

Prostate cancer is the most common non-cutaneous cancer among American men, disproportionately affecting Black men more than their White counterparts. Black men are diagnosed at a younger age, exhibit more advanced disease stages, and have less access to treatment, leading to worse clinical outcomes. However, according to the most recent 2018 recommendations by the United States Preventive Services Task Force, routine prostate-specific antigen (PSA) screening is typically not recommended for men aged 40 to 54 due to potential overtreatment concerns arising from excessive screening in this low-risk group \citep{rawla2019epidemiology, grossman2018screening, barocas2010racial, hoffman2001racial}.

\begin{figure}[ht]
    \centering
        \begin{subfigure}[t]{0.49\columnwidth}
            \includegraphics[width=\linewidth]
            {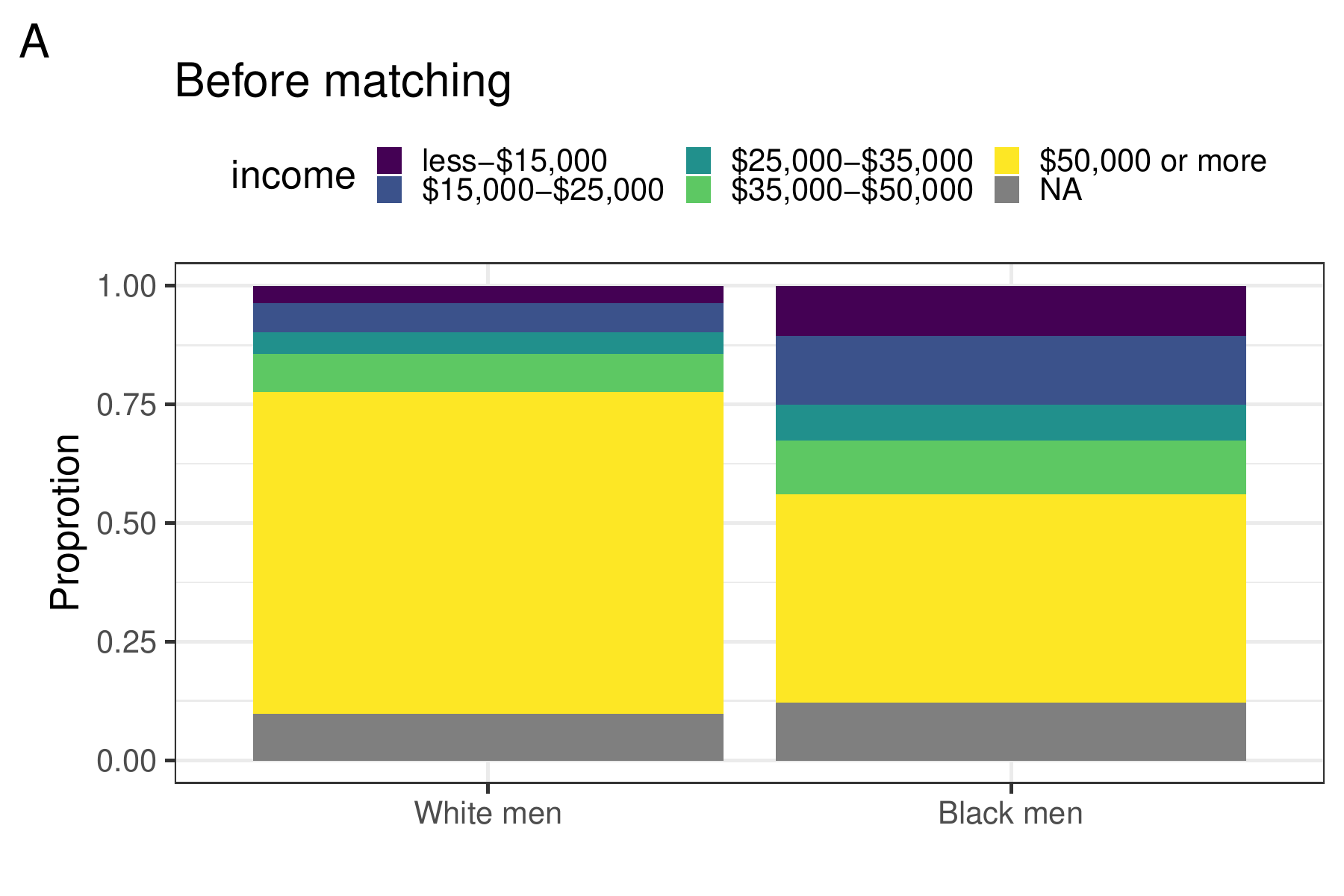}
        \end{subfigure}
        \hfill
        \begin{subfigure}[t]{0.49\columnwidth}
            \includegraphics[width=\linewidth]{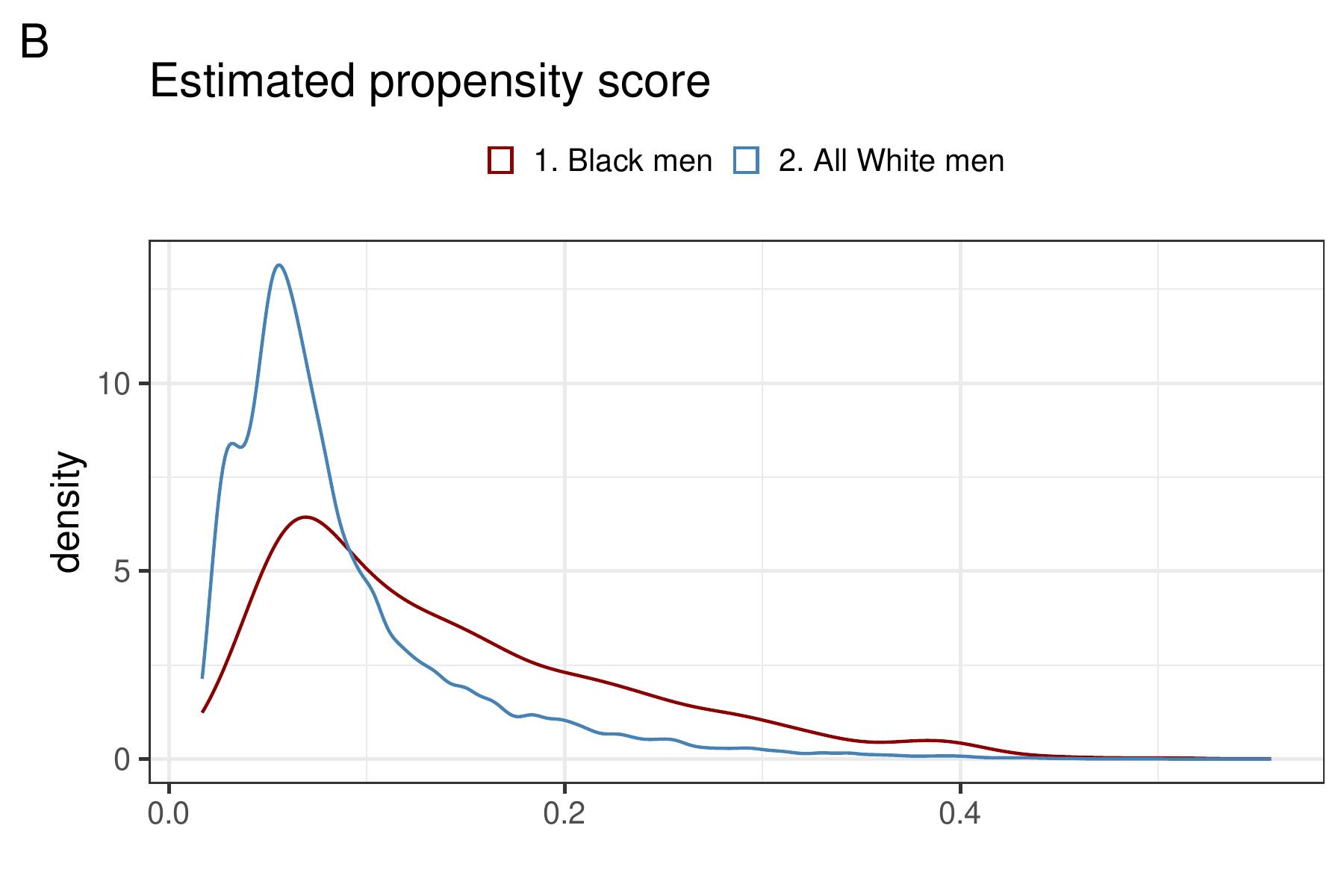}
        \end{subfigure}
    \caption{\textbf{Panel A: Illustration of before matching covariate: income imbalance between the Black and White men.} Stacked bar graphs of income for Black and White men before matching. Before matching a higher proportion of white men had annual income above \$50,000 while a lower proportion of white men with annual income less than \$15,000 compared to Black men.
    \textbf{Panel B: Illustration of before matching covariates imbalance between the Black and White men.} Estimated propensity score distribution of Black and White men. Before matching the marginal covariates distribution of Black men is different from the distribution of White men.
    }
    \label{fig:bef_match}
\end{figure}

There are numerous disparities between younger Black and White men. As depicted in Panel A of Figure \ref{fig:bef_match} and the initial three columns of Table \ref{tbl: table1}, 2020 BRFSS data shows that among the 23,344 White and 2,507 Black younger men recorded, a higher proportion of White men earned over \$50,000 annually (67.9\% vs. 44.0\%). Additionally, fewer White men earned under \$15,000 annually (3.7\%) compared to Black men (10.5\%). White men also exhibited higher rates of college or technical college graduation, lower rates of BMI above 30, more widespread health insurance coverage, better self-reported health, and higher marriage rates. The age during the survey was comparable for both groups. Panel B in Figure \ref{fig:bef_match} further exhibits the overlap of the distributions of \citeauthor{rosenbaum1983central}'s \citeyearpar{rosenbaum1983central} propensity score among white and black men. The propensity score traditionally refers to the conditional probability of treatment assignment given the observed covariates \citep{rosenbaum1983central}. In health disparities research, propensity score refers to the conditional probability of a study participant being a certain race/ethnicity (see, e.g., \citealp{nogueira2022association}), and it is still a useful statistical quantity that provides a scalar summary of the extent to which the two comparison groups differ in observed covariates used to estimate the propensity score. 

Differences in these observed characteristics between Black and White populations often prompt researchers to control for selected observed covariates. Using the BRFSS data, \citet{qian2023investigating} found that the 2-year PSA screening rate of White young men in 2020 was significantly lower than that among Black young men after adjusting for covariates including age at the time of the survey, education level, annual income, insurance coverage, marital status, smoking status, body mass index, self-reported overall status of health, and an indicator of having a personal doctor in a regression-based analysis (odds ratio: 0.69; 95\% CI: [0.54, 0.87]). 

\subsection{Disentangling the contributions from various factors to the disparity gap}

Our goal is to use the proposed algorithm in Section \ref{subsec: tripartite in disparity research} to create nested matched comparisons and investigate the contribution of various modifiable factors, such as income, insurance, and other socioeconomic factors, to the observed disparities. We examined the BRFSS survey data in 2020. The study population consists of non-Hispanic Black and White men aged 40 to 54 who did not have a previous history of prostate cancer, and the primary outcome of interest is the receipt of PSA screening within two years prior to the BRFSS survey.


Table \ref{tbl: table1} shows the covariate balance among the $2,507$ Black men, $24,344$ White men before matching, and three matched comparison groups. Design \textsf{M1} constructed a matched comparison group consisting of White men who closely resemble the Black men in biology-related variables, including age, BMI, health status and smoking status, while remaining similar to the white population before matching in other factors including income, health-care provider status, education level, health insurance status and marital status. The matched comparison group in Design \textsf{M2}, in addition to resembling the Black men in all biology-related variables as in \textsf{M1}, also resembles the Black men in income. Lastly, the White comparison group in design \textsf{M3} resembles the White men in all covariates considered in the study. The zigzag line that cuts Table \ref{tbl: table1} helps visualize which covariates are closely matched with Black men and which covariates are kept close to the original White population: variables above the line are designed to resemble the Black men while those below the line are designed to resemble the White men before matching. The balance in each covariate is examined by calculating the absolute standardized mean differences (SMDs) between the comparison group and the Black men group. An SMD less than $0.1$ is typically considered a good balance in empirical health services research \citep{silber2001multivariate}. To summarize, all three designs, $\textsf{M1}$ through $\textsf{M3}$, now largely resemble the Black group in age, BMI, health status, and smoking status. The design $\textsf{M1}$ further resembles the White men before matching in income, HCP status, education, health insurance, and marital status. Design $\textsf{M2}$ selects White men who are similar to those selected in $\textsf{M1}$ in all but income: White comparison group in $\textsf{M1}$ have their income distribution remaining at the level of White men before matching, while those in $\textsf{M2}$ now resemble the Black men in income.

\begin{table}[ht]
\caption{Covariate balance table in PSA screening, racial disparity study before matching and three matched samples (M1, M2 and M3) constructed using the proposed algorithm. The zig-zag line divides the presence of covariates within each match. }
\centering
\begin{threeparttable}[h]
\scalebox{0.7}{
\begin{tabular}{r|ccc|cccccc}
  \hline
 & White & Black & SMD & White & SMD & White & SMD & White & SMD \\ 
 & (Bef) &  & (Bef) & (M1) & (M1) & (M2) & (M2) & (M3) & (M3) \\
  \hline
\textbf{Sample Size} & 24,344 & 2,507 &  & 2,507 &  & 2,507 &  & 2,507 &  \\ 
  \textbf{AGE (\%)} &   &   &  &   &  &   &  &   &  \\ 
  40-44 &   7442 (30.6)  &   755 (30.1)  & 0.011 &   756 (30.2)  & 0.002 &   755 (30.1)  & 0 &   753 (30.0)  & 0.002 \\ 
  45-49 &   7704 (31.6)  &   812 (32.4)  & 0.017 &   811 (32.3)  & 0.002 &   811 (32.3)  & 0.002 &   784 (31.3)  & 0.024 \\ 
  50-54 &   9198 (37.8)  &   940 (37.5)  & 0.006 &   940 (37.5)  & 0 &   941 (37.5)  & 0 &   970 (38.7)  & 0.025 \\ 
  \textbf{BMI (\%)} &   &   &  &   &  &   &  &   &  \\ 
  $<$25 &   4605 (18.9)  &   471 (18.8)  & 0.003 &   470 (18.7)  & 0.003 &   466 (18.6)  & 0.005 &   457 (18.2)  & 0.015 \\ 
  25-30 &   9879 (40.6)  &   903 (36.0)  & 0.095 &   904 (36.1)  & 0.002 &   908 (36.2)  & 0.004 &   906 (36.1)  & 0.002 \\ 
  $>$30 &   9291 (38.2)  &  1034 (41.2)  & 0.061 &  1033 (41.2)  & 0 &  1042 (41.6)  & 0.008 &  1068 (42.6)  & 0.029 \\ 
  Missing &    569 ( 2.3)  &    99 ( 3.9)  & 0.092 &   100 ( 4.0)  & 0.006 &    91 ( 3.6)  & 0.017 &    76 ( 3.0)  & 0.052 \\ 
  \textbf{Health status (\%)} &   &   &  &   &  &   &  &   &  \\ 
  Excellent &   5836 (24.0)  &   522 (20.8)  & 0.077 &   522 (20.8)  & 0 &   525 (20.9)  & 0.002 &   492 (19.6)  & 0.029 \\ 
  Very good &   9522 (39.1)  &   712 (28.4)  & 0.228 &   712 (28.4)  & 0 &   707 (28.2)  & 0.004 &   719 (28.7)  & 0.006 \\ 
  Good &   6609 (27.1)  &   832 (33.2)  & 0.133 &   832 (33.2)  & 0 &   837 (33.4)  & 0.004 &   891 (35.5)  & 0.05 \\ 
  Fair &   1750 ( 7.2)  &   330 (13.2)  & 0.199 &   330 (13.2)  & 0 &   332 (13.2)  & 0 &   314 (12.5)  & 0.023 \\ 
  Poor &    591 ( 2.4)  &    98 ( 3.9)  & 0.086 &    99 ( 3.9)  & 0 &    95 ( 3.8)  & 0.006 &    88 ( 3.5)  & 0.023 \\ 
  Missing &     36 ( 0.1)  &    13 ( 0.5)  & 0.073 &    12 ( 0.5)  & 0 &    11 ( 0.4)  & 0.018 &     3 ( 0.1)  & 0.073 \\ 
  \textbf{Smoking status (\%)} &   &   &  &   &  &   &  &   &  \\ 
  Never Smoker &  13038 (53.6)  &  1543 (61.5)  & 0.16 &  1543 (61.5)  & 0 &  1543 (61.5)  & 0 &  1552 (61.9)  & 0.008 \\ 
  Former Smoker &   6667 (27.4)  &   385 (15.4)  & 0.296 &   384 (15.3)  & 0.002 &   382 (15.2)  & 0.005 &   373 (14.9)  & 0.012 \\ 
  Current Smoker &   4505 (18.5)  &   560 (22.3)  & 0.094 &   562 (22.4)  & 0.002 &   569 (22.7)  & 0.01 &   575 (22.9)  & 0.015 \\ 
  Missing &    134 ( 0.6)  &    19 ( 0.8)  & 0.024 &    18 ( 0.7)  & 0.012 &    13 ( 0.5)  & 0.036 &     7 ( 0.3)  & 0.06 \\ \cline{5-6}
  \textbf{Income (\%)} &   &   &  &   &  & \multicolumn{1}{|l}{}  &  &   &  \\ 
  less-\$15,000 &    899 ( 3.7)  &   264 (10.5)  & 0.267 &    92 ( 3.7)  & 0.267 &   \multicolumn{1}{|l}{266 (10.6)}  & 0.004 &   254 (10.1)  & 0.016 \\ 
  \$15,000-\$25,000 &   1491 ( 6.1)  &   363 (14.5)  & 0.279 &   156 ( 6.2)  & 0.276 &   \multicolumn{1}{|l}{364 (14.5)}  & 0 &   369 (14.7)  & 0.007 \\ 
  \$25,000-\$35,000 &   1084 ( 4.5)  &   189 ( 7.5)  & 0.127 &   118 ( 4.7)  & 0.118 &   \multicolumn{1}{|l}{186 ( 7.4)}  & 0.004 &   182 ( 7.3)  & 0.008 \\ 
  \$35,000-\$50,000 &   1959 ( 8.0)  &   284 (11.3)  & 0.112 &   239 ( 9.5)  & 0.061 &   \multicolumn{1}{|l}{283 (11.3)}  & 0 &   295 (11.8)  & 0.017 \\ 
  \$50,000 or more &  16538 (67.9)  &  1104 (44.0)  & 0.496 &  1653 (65.9)  & 0.454 &  \multicolumn{1}{|l}{1106 (44.1)}  & 0.002 &  1109 (44.2)  & 0.004 \\ 
  Missing &   2373 ( 9.7)  &   303 (12.1)  & 0.077 &   249 ( 9.9)  & 0.071 &   \multicolumn{1}{|l}{302 (12.0)}  & 0.003 &   298 (11.9)  & 0.006 \\ \cline{7-8}
  \textbf{Has HCP (\%)} &   &   &  &   &  &   &  & \multicolumn{1}{|l}{}   &  \\ 
  No &   5889 (24.2)  &   637 (25.4)  & 0.028 &   599 (23.9)  & 0.035 &   599 (23.9)  & 0.035 &   \multicolumn{1}{|l}{576 (23.0)}  & 0.056 \\ 
  Yes &  18431 (75.7)  &  1867 (74.5)  & 0.028 &  1907 (76.1)  & 0.037 &  1907 (76.1)  & 0.037 &  \multicolumn{1}{|l}{1931 (77.0)}  & 0.058 \\ 
  Missing &     24 ( 0.1)  &     3 ( 0.1)  & 0 &     1 ( 0.0)  & 0.032 &     1 ( 0.0)  & 0.032 &    \multicolumn{1}{|l}{0 ( 0.0)}  & 0.032 \\ 
  \textbf{Level of education (\%)} &   &   &  &   &  &   &  & \multicolumn{1}{|l}{}   &  \\ 
  not finished high school &   1070 ( 4.4)  &   194 ( 7.7)  & 0.139 &   111 ( 4.4)  & 0.139 &   111 ( 4.4)  & 0.139 &   \multicolumn{1}{|l}{194 ( 7.7)}  & 0 \\ 
  graduated High School &   6294 (25.9)  &   805 (32.1)  & 0.137 &   680 (27.1)  & 0.11 &   680 (27.1)  & 0.11 &   \multicolumn{1}{|l}{828 (33.0)}  & 0.02 \\ 
  College/Tech &   6226 (25.6)  &   717 (28.6)  & 0.068 &   626 (25.0)  & 0.081 &   626 (25.0)  & 0.081 &  \multicolumn{1}{|l}{711 (28.4)}  & 0.005 \\ 
  graduated College/Tech &  10696 (43.9)  &   781 (31.2)  & 0.265 &  1084 (43.2)  & 0.25 &  1084 (43.2)  & 0.25 &   \multicolumn{1}{|l}{767 (30.6)}  & 0.012 \\ 
  Missing &     58 ( 0.2)  &    10 ( 0.4)  & 0.037 &     6 ( 0.2)  & 0.037 &     6 ( 0.2)  &    0.037 &   \multicolumn{1}{|l}{7 ( 0.3)}  & 0.018 \\ 
  \textbf{Has health insurance (\%)} &   &   &  &   &  &   &  & \multicolumn{1}{|l}{}  &  \\ 
  No &   2424 (10.0)  &   400 (16.0)  & 0.179 &   247 ( 9.9)  & 0.182 &   247 ( 9.9)  & 0.182 &   \multicolumn{1}{|l}{325 (13.0)}  & 0.09 \\ 
  Yes &  21900 (90.0)  &  2102 (83.8)  & 0.185 &  2257 (90.0)  & 0.185 &  2257 (90.0)  & 0.185 &  \multicolumn{1}{|l}{2182 (87.0)}  & 0.095 \\ 
  Missing &     20 ( 0.1)  &     5 ( 0.2)  & 0.026 &     3 ( 0.1)  & 0.026 &     3 ( 0.1)  & 0.026 &  \multicolumn{1}{|l}{0 ( 0.0)}  & 0.052 \\ 
  \textbf{Marital status (\%)} &   &   &  &   &  &   &  & \multicolumn{1}{|l}{}  &  \\ 
  Never Married &   2947 (12.1)  &   702 (28.0)  & 0.405 &   327 (13.0)  & 0.382 &   327 (13.0)  & 0.382 &   \multicolumn{1}{|l}{707 (28.2)}  & 0.005 \\ 
  Married/couple &  17095 (70.2)  &  1206 (48.1)  & 0.461 &  1727 (68.9)  & 0.434 &  1727 (68.9)  & 0.434 &  \multicolumn{1}{|l}{1209 (48.2)}  & 0.002 \\ 
  Other &   4191 (17.2)  &   581 (23.2)  & 0.15 &   440 (17.6)  & 0.14 &   440 (17.6)  & 0.14 &   \multicolumn{1}{|l}{584 (23.3)}  & 0.002 \\ 
  Missing &    111 ( 0.5)  &    18 ( 0.7)  & 0.026 &    13 ( 0.5)  & 0.026 &    13 ( 0.5)  & 0.026 &     \multicolumn{1}{|l}{7 ( 0.3)}  & 0.052 \\ 
   \hline
\end{tabular}
}
\begin{tablenotes}
       \item \textsf{BMI = body mass index; HCP = health-care provider.}
\end{tablenotes}

\end{threeparttable}\label{tbl: table1}
\end{table}

\subsection{Outcome analysis}

Table \ref{tbl: outcome} displays the PSA screening rates, estimated odds ratios, and associated $95\%$ confidence intervals and p-values between the White and Black men based on the R package \textsf{exact2x2} in various comparisons. Initially, the unmatched data showed a screening rate of $13.0\%$ for White men and $15.6\%$ for Black men. In the first matched comparison that controls for biology-related variables, the screening rate among the white men in the comparison group decreases moderately from $13.0\%$ to $12.6\%$ and was significantly lower compared to that in the Black men (odds ratio: 0.77; 95\% CI: [0.65, 0.91]). In a second comparison that further matched on income, the screening rate among the white men in the comparison group further decreased to $12.0\%$ and the gap between the two groups widened (odds ratio: 0.72; 95\% CI: [0.61, 0.86]). In the last design that matched on all variables, the screening rate among the white men decreased to $11.0\%$ and the gap further widened (odds ratio: 0.65; 95\% CI: [0.55, 0.78]). 

In this analysis, a na\"ive, unadjusted analysis would mask a substantial difference among White and Black men because the subgroup of white men who had a lower income and in general at a less advantageous socioeconomic status tended to have lower PSA screening rate. A comparison between the results obtained from $\textsf{M1}$ and $\textsf{M2}$ seems to suggest that income alone only had a moderate impact on the PSA screening rate. A comparison between $\textsf{M1}$ and $\textsf{M3}$ suggests that all socioeconomic status (SES) factors put together, including income, whether a person has a personal doctor, level of education, health insurance coverage and marital status, had a large impact on the PSA screening rate and widen the gap between White and Black men from 12.6\% - 15.6\% = -3.0\% to 11.0\% - 15.6\% = -4.6\%. 
Our nested-matched case study demonstrates a disparity in PSA screening practices, with younger Black men undergoing more screenings than their White counterparts, counter to current United States Preventive Services Taskforce (USPSTF) guidelines. This over-screening could potentially lead to unnecessary treatments and associated harm. This issue becomes even more severe considering the socioeconomic disparities we identified. After matching these characteristics, it was found that younger Black men, who typically have less access to healthcare resources, are being over-screened, thereby subjecting them to potential harm from unnecessary interventions. Simultaneously, younger Black men are at a higher risk for aggressive prostate cancer, which requires a careful balance between over-screening and early detection. Hence, our findings underscore the need for developing individualized, evidence-based screening strategies that account for both medical risk factors and social determinants of health. This can lead to more accurate screening, improved patient outcomes, and a reduction in health disparities.


\begin{table}[ht]
\centering
\caption{PSA screening rate and odds ratios for prostate-specific antigen screening in young White and Black men(2020). }\label{tbl: outcome}
\begin{tabular}{ccccl}
\hline
\textbf{Comparison}  &  
\multicolumn{2}{c}{\textbf{PSA Screening Rate}} & \textbf{Odds Ratio (OR)} & \textbf{P-value} \\ 
 & White & Black &(95\% CI) & \\
\hline

\textsf{Unmatched}   &    0.130      & 0.156       &                   &    \\
\textsf{M1}  &    0.126        & 0.156     & 0.770(0.651, 0.909)                     &  $<.001$ \\ 
\textsf{M2}  &    0.120       & 0.156       & 0.724(0.610, 0.859)                     &  $<.001$  \\ 
\textsf{M3}  &    0.110       & 0.156      & 0.650(0.545, 0.775)                     & $<.001$    \\  
\hline
\end{tabular}

\end{table} 

\subsection{Data availability}
Data supporting the analysis this article is available at: \url{https://www.cdc.gov/brfss/annual\_data/annual\_data.html}
\section{Software}\label{sec: code}

\subsection{A basic match with minimal input}\label{subsec: minimal match}

\begin{lstlisting}[language=R, float, caption = Basic match, label={lst: minimal}]
# main code ---------------------------------------------------------------
rm(list = ls())
library(match2C)
library(tidyverse)
## data preparation ----
## load data
load("./data/minimal_data.rdata")

## Fit a propensity score model
### design matrix covariate names
var_list_desgin <-
  setdiff(grep(paste(var_list, collapse = "|"),
               names(min_dat),
               value = T), var_list)
### propensity calculation formula
propen_f <- as.formula(paste("Z~",
                             paste(var_list_desgin, collapse = ' + ')))

propensity = glm(propen_f, family = binomial, data = min_dat)$fitted.values
min_dat$propensity = propensity

## use `match2C` function ----
## constructing the optimal match pairs
X <- min_dat %>% select(!!var_list_desgin)

matching_output_minial = match_2C(
  Z = min_dat$Z,
  X = X,
  propensity = min_dat$propensity,
  caliper_left = 0.5,
  caliper_right = 0.5,
  k_left = 500,
  k_right = 500,
  dataset = min_dat
)

# Check feasibility 
matching_output_minial$feasible

# Check data frame organized in matched set indices 
head(matching_output_minial$matched_data_in_order, 6)
\end{lstlisting}

We discuss with examples basic functionalities in the \textsf{R} package \textsf{match2C} and showcase how to implement the method proposed in Section \ref{sec: code} within the \textsf{match2C} infrastructure. We start by importing the package and selecting a subset of the dataset used in Section \ref{sec: case study} for a one-step optimal match. The code snippet labeled Listing \ref{lst: minimal} demonstrates this process. To simplify the explanation, we will focus on three categorical variables: age, BMI, and health status. The treatment variable represents the patient's race, where $1$ indicates the subject is black and $0$ indicates the subject is white. The outcome variable indicates the patient's PSA screening status. We convert all three categorical variables into dummy variables based on their levels and combine them as covariates ($\mathbf{X}$) and treatment (Z) for matching. We then fit a propensity score model.

The function \textsf{match\_2C} is a modified version of \textsf{match\_2C\_list} with predefined distance structures. The left network uses a Mahalanobis distance between covariates $\mathbf{X}$, while the right network uses a L1 distance between the estimated propensity score. A large penalty is imposed to prioritize balancing the propensity score distributions in the treated and matched control groups. This is followed by minimizing the sum of Mahalanobis distances within matched pairs. \textsf{match\_2C} also enables fine-balancing the joint distribution of a few important covariates. A typical hierarchical structure is as follows: fine-balance $>>$ propensity score distribution $>>$ within-pair Mahalanobis distance.

The function \textsf{match\_2C} also allows users to specify two caliper sizes for the propensity scores: \textsf{caliper\_left} for the left network and \textsf{caliper\_right} for the right network. If a caliper other than the propensity score caliper or an asymmetric caliper is desired \citep{yu2020matching}, then the function \textsf{match\_2C\_list} function can entertain these options. 
Additionally, users have the option to limit the number of edges connected to each treated unit using parameters \textsf{k\_left} and \textsf{k\_right}. By default, each treated subject in the network is connected to all \textsf{n\_c} control subjects. However, setting \textsf{k\_left} and \textsf{k\_right} allows users to connect each treated subject only to the \textsf{k\_left} or \textsf{k\_right} control subjects closest in the propensity score. For example, setting \textsf{k\_left} = 500 means that each treated subject will be connected to a maximum of 500 control subjects with the closest propensity score in the left network. Similarly, the \textsf{k\_right} parameter would connect each treated subject to the closest \textsf{k\_right} controls in the right network. 
\textsf{caliper\_low}, \textsf{caliper\_high}, \textsf{k\_left} and \textsf{k\_right} options can be used in combination. Our simple example in Listing \ref{lst: minimal} demonstrates the usage of calipers on both the left and right networks while limiting the maximum number of edges connected to each treated subject at the same time. 
In some applications, if the caliper sizes are set too small, it may be impossible to find a feasible match. In such cases, it is advised to increase the caliper size and/or relax the constraints for the matching.

\subsection{Constructing a customized tripartite network}\label{subsec: nested match code}

We next describe how to implement our proposed method mentioned in Section \ref{subsec: tripartite in disparity research} using the \textsf{match2C} package.
In the above one-step example, for instance, suppose one aims to create a matched comparison group of white men that closely resembles black men in Age and BMI covariates, while maintaining similarity to the white population before matching in terms of health status. 

To do this we use the \textsf{create\_list\_from\_scratch} command to derive a list representation of a treatment-by-control distance matrix consisting of the following arguments:
i) start\_n: a vector containing the node numbers of the start nodes of each arc in the network. ii) end\_n: a vector containing the node numbers of the end nodes of each arc in the network. iii) d: a vector containing the integer cost of each edge in the network. Nodes $1$, $2$,...,$n_t$ correspond to $n_t$ treatment nodes, and $n_t + 1$, $n_t + 2$, ..., $n_t + n_c$ correspond to $n_c$ control nodes. Note that start\_n, end\_n, and d have the same lengths, all of which are equal to the number of edges. Besides, this function allows users to construct a (possibly sparse) distance list (arguments \textsf{k} and \textsf{caliper\_low}) with a possibly user-specified distance measure (argument \textsf{method}). 

In our example, we first derived two sets of propensity scores calculated based on different covariate groups, i.e., one based on Age and BMI, the other only based on health status. To construct the left distance list shown in code chunk Listing \ref{lst: left_dist_list}, we compute robust Mahalanobis distances between black and white patients based on Age and BMI alone. We also specified the parameter \textsf{caliper\_low}  to only include distances between black and white patients within a $0.5$ propensity score, and we also specified the parameter \textsf{k} so that each black man is connected to the closest $500$ white men in the propensity score.
\begin{lstlisting}[language=R, caption = Compute the left distance list, label={lst: left_dist_list}]

## Compute left distance list
dist_list_left = create_list_from_scratch(
  Z = min_dat$Z,
  X = min_dat %>% select(!!var_list_left_desgin),
  p = left_propensity,
  caliper_low = 0.5,
  k = 500,
  method = 'robust maha'
)

\end{lstlisting}

To construct the right distance list shown in code chunk Listing \ref{lst: right_dist_list}, we compute robust Mahalanobis distances based on the health status between a sampled subgroup of white patients and the entire white patient population. And similar to the left distance list, we set the same \textsf{k} and \textsf{caliper\_low} arguments to sparsify the distance list structure.
\begin{lstlisting}[language=R, caption = Compute the right distance list, label={lst: right_dist_list}]

## Compute right distance list

### sample n_t white patients
set.seed(100)
pind_sampled <-
  min_dat %>% filter(Z == 0) %>%
  select(pind) %>% unlist %>% sample(., n_t)

### set Z_right identifying sampled subgroup and original population 
Z_right <- c(rep(1, n_t), rep(0, n_c))
X <- min_dat %>% filter(Z == 0) %>%
  select(!!var_list_right_desgin, pind, right_propensity)
X_right <- X %>% filter(pind %in% pind_sampled) %>% rbind(., X)

### compute the distance
dist_list_right = create_list_from_scratch(
  Z = Z_right,
  X = X_right %>% select(-pind,-right_propensity),
  p = X_right %>% select(right_propensity) %>% unlist,
  caliper_low = 0.5,
  k = 500,
  method = 'robust maha'
)

\end{lstlisting}

Once we have constructed two distance structures, they are passed into the function \textsf{match\_2C\_list} directly. This function is similar to \textsf{match\_2C}, with the distinction that it requires at least one distance list as input. We further specify a penalty of $10$ so that the algorithm prioritizes balancing the health status distribution in the matched white patients and the white population before matching, followed by minimizing the total within-matched-pair robust Mahalanobis distances between the black men and matched white men shown in code chunk Listing \ref{lst: match2C_list}. Objects returned by the family of matching functions \textsf{match\_2C}, \textsf{match\_2C\_list}, and \textsf{match\_2C\_mat} are the same in the format: a list of the following three elements: i) \textsf{feasible}: 0/1 depending on the feasibility of the matching problem; ii) \textsf{data\_with\_matched\_set\_ind}: a data frame that is the same as the original data frame, except that a column called matched\_set and a column called distance are added to it. Variable \textsf{matched\_set} assigns $1$,$2$,...,$n_t$ to each matched set, and \textsf{NA} to controls not matched to any treated. Variable \textsf{distance} records the control-to-treated distance in each matched pair and assigns \textsf{NA} to all treated and controls that are left unmatched. If there is no feasible match, \textsf{NULL} will be returned; iii) \textsf{matched\_data\_in\_order}: a data frame organized in the order of matched sets and otherwise the same as \textsf{data\_with\_matched\_set\_ind}. Null will be returned if the matching is unfeasible.

\begin{lstlisting}[language=R, caption = Derive the match comparisons, label={lst: match2C_list}]

matching_output = match_2C_list(
  Z = min_dat$Z,
  dataset = min_dat,
  dist_list_1 = dist_list_left,
  dist_list_2 = dist_list_right,
  lambda = 10,
  controls = 1
)

\end{lstlisting}

\subsection{Balance assessment}

The code chunk Listing \ref{lst: check_balance} showcases the function \textsf{check\_balance}, which serves as a tool to assess and visualize balance between different groups of patients after performing matching. The function \textsf{check\_balance} takes an indicator of the black patient \textsf{Z}, an object returned by \textsf{match\_2C\_list} (or \textsf{match\_2C} or \textsf{match\_2C\_mat}), a vector of covariate names for which balance needs to be checked as inputs. Note function \textsf{check\_balance} could also plot the distribution of the propensity score among the treated subjects (black patients in this example), all control subjects (white patients in this example), and the matched control subjects by setting option \textsf{plot\_propens = TRUE}. Estimated propensity scores should be provided as input using the \textsf{propens} option.

\begin{lstlisting}[language=R, caption = Balance assessment, label={lst: check_balance}]

tb_example = check_balance(
    Z, 
    matching_output, 
    cov_list = c(var_list_left_desgin, var_list_right_desgin), 
    plot_propens = FALSE
)

print(tb_example)

\end{lstlisting}
\section{Discussion}
\label{sec: discussion}
In this article, we proposed a novel matching algorithm for health disparities research. Our method builds upon the idea of tapered matching but deviates from it by constructing matched comparison groups that not only resemble the target ``treated" group (e.g., black men) in many covariates, but also remain identical to the source population (e.g., white men) in other covariates. We successfully applied our proposed method to investigating the impact of modifiable factors on health disparities in PSA screening rates among non-Hispanic Black and White men aged $40$ to $54$ with no history of prostate cancer. The analysis, based on 2020 BRFSS survey data, yielded important insights. While income alone exhibited a moderate impact on PSA screening rates, considering all socioeconomic status (SES) factors together—such as income, access to a personal doctor, level of education, health insurance coverage, and marital status—revealed a substantial influence on the PSA screening rate. The combined effect significantly widened the PSA screening rate gap between white and black men patients from $-3.0\%$ to $-4.6\%$. These findings emphasize the significance of identifying and considering more modifiable determinants of socioeconomic status beyond solely focusing on a few SES variables when assessing disparities in PSA screening rates. By acknowledging the broader range of factors, such as income, access to healthcare, education, health insurance, and marital status, we gain a more comprehensive understanding of the complex interplay between socioeconomic conditions and PSA screening rates. This broader perspective enables us to develop targeted interventions and policies aimed at reducing health disparities and promoting equitable access to preventive healthcare services.



However, despite the advantage of network flow optimization in finding the global minimum covariate distance between treated and matched comparisons, conducting such optimization in some large administrative datasets (e.g., the BRFSS database in the case study) can become computationally challenging in dense networks, characterized by two disjoint sets of nodes where each node in the first set is connected to every node in the second set.
For example, solving a bipartite network flow optimization problem that optimally pairs $T$ = 2,507 black male patients with 2,507 white male patients from a pool of $C$ = 23,344 white male patients would require a time complexity of order $O(T^3)$\citep{yu2020matching}. 
\citet{yu2020matching} proposed an approach to enhance the dense matching efficiency and sparse matching feasibility with an iterative variant of Glover's algorithm \citep{glover1967maximum, lipski1981efficient}. This approach uses binary search to find the minimum caliper size that satisfies pair-matching feasibility in a doubly convex bipartite graph with time complexity in the order of $O(T)$. As a result, this approach significantly reduced the pool of potential matches, improving efficiency in constructing optimally matched comparisons with the time complexity of order $O(T^2 \log T)$. Additionally, this approach mitigated infeasibility risk, where a control subject could not be matched with a treated subject.  In the tripartite network flow optimization scenarios, it is of interest to address whether applying \citet{yu2020matching}'s sparse matching feasibility check algorithm twice, once on the left bipartite network and once on the right bipartite network, would find the minimum caliper size for the entire tripartite network. As combining the two local minimum caliper sizes may not necessarily ensure a global minimum caliper size.

Indeed, the utilization of an ad-hoc, fixed caliper size in the case study presents some limitations, particularly concerning its data-dependent nature. While it may be feasible and efficient for the specific BRFSS 2020 dataset used in the study, this fixed caliper size approach could be further improved to adapt to the new dataset.
For example, in an optimal pair-match problem, when 95\% estimated propensity scores for treated subjects are within the range of $0.5$ to $0.6$, it is easy to pair the control subjects to the treated subjects in this range, supposing a fixed minimum caliper size as $0.1$, as finding a paired control subject between the propensity score $0.4$ to $0.7$ is straightforward by the definition of the estimated propensity score. More precisely, the optimization problem will still be feasible if the caliper size is increased to $0.2$ for pairing a control for this portion of treated subjects. 
However, the few treated subjects with estimated propensity scores around $0.8$ may be merely able to find a paired comparison when the caliper size is $0.1$. 
In other words, it is essential to tailor and refine the fixed minimum caliper size to suit the estimated propensity score of each treated subject. As a result, there is a pressing requirement to develop adaptive methods capable of optimizing the feasibility check time complexity in such scenarios.

\clearpage
\bibliographystyle{apalike}
\bibliography{sample}

\end{document}